# 783-MHz fundamental repetition rate all-fiber ring laser mode-locked by carbon nanotubes

Maolin Dai[1,2], Bowen Liu[1,2], Yifan Ma[1,2], Takuma Shirahata[1,2], Ruoao Yang[3], Zhigang Zhang[3], Sze Yun Set[1,2,*], and Shinji Yamashita[1,2]

[1]*Department of Electrical Engineering and Information Systems, The University of Tokyo, Bunkyo-ku, Tokyo 113-8656, Japan*
[2]*Research Center for Advanced Science and Technology, The University of Tokyo, Meguro-ku, Tokyo 153-8904, Japan*
[3]*State Key Laboratory of Advanced Optical Communication Systems and Networks, School of Electronics, Peking University, Beijing 100871, China*

E-mail: maolin@cntp.t.u-tokyo.ac.jp; set@cntp.t.u-tokyo.ac.jp

Abstract:
We demonstrate a 783-MHz fundamental repetition rate mode-locked Er-doped all-fiber ring laser with a pulse width of 623 fs. By using carbon nanotubes (CNT) saturable absorber (SA), a relatively low self-starting pump threshold of 108 mW is achieved. The laser has a very compact footprint less than 10 cm × 10 cm, benefiting from the all-active-fiber cavity design. The robust mode-locking is confirmed by the low relative intensity noise (RIN) and a long-term stability test. We propose a new scheme for generating high repetition rate femtosecond optical pulses from a compact and stable all-active-fiber ring oscillator.

Owing to the widespread applications in both scientific research and industries such as frequency comb-based spectroscopy [1-3], laser processing [4, 5], and bio-imaging [6], high repetition rate fiber lasers have been developed intensively and many breakthroughs have been achieved. Techniques such as active mode-locking [7], passively harmonic mode-locking [8, 9], dissipative four-wave mixing [10] and mode filtering [11] could achieve a very high repetition rate, however, they exhibit a higher instability in terms of output performance compared to fundamentally passive mode-locking [12]. Previous works have shown that ultrahigh repetition rates can be obtained by using short fiber Fabry-Perot resonators [13-15]. Ring lasers can get rid of standing waves and spatial hole burning effect that always happen in linear lasers, however, it is challenging for realizing high repetition rate ring lasers.

Recently, high repetition rates have been successfully achieved from Yb-doped fiber ring lasers [16-19]. Compared to Yb-doped silica fibers, Er-doped silica fibers have lower gain efficiency, therefore, many of the reported high repetition rate fiber lasers in 1.5-μm band are based on highly Er-doped or Er/Yb co-doped phosphate glass fibers [20-22]. In 2016, J. Zhang et al. demonstrated a 517-MHz ring fiber laser mode-locked by nonlinear polarization evolution (NPE) using commercialized highly Er-doped silica fiber [23]. In 2023, using the same oscillator design, the cavity length was further decreased to 22.5 cm, corresponding to a repetition rate of 895 MHz [24]. However, those lasers contain the free space components which require fine alignment. All-fiber configuration would better benefit the compactness and robustness of the high repetition rate fiber lasers. In 2017, 384 MHz repetition rate was achieved by an all-fiber NPE ring laser using silica gain fiber [25]. However, all-fiber NPE lasers with higher repetition rate are hard to achieve owing to the limited space for polarization management.

Reports have shown that using slow saturable absorber (SA) such as carbon nanotubes (CNT) can realize high repetition rate mode-locking with femtosecond pulse width [26-28]. Benefiting from its low saturation intensity, self-started femtosecond pulses could be achieved in all-fiber laser cavities under a relatively low pump power. By using CNT-SA, high repetition rates of over 300 MHz have been realized [26-28]. We summarize the development of Er-doped fiber ring lasers delivering high repetition rates as Table. I.

Here, we report an Er-doped all-fiber ring laser mode-locked by CNT-SA that delivers 783-MHz fundamental repetition rate and 623-fs pulse width at the center wavelength of 1562 nm. The laser oscillator has a very compact size with footprint of less than 10 cm × 10 cm, where only 24-cm Er-doped silica fiber, an all-in-one integrated device and CNT-SA coated on fiber ferrule are used in the laser cavity, making it a robust all-active-fiber configuration. Turn-key mode-locking operation can be achieved under a low self-starting pump threshold of 108 mW. To the best of our knowledge, 783 MHz is the highest fundamental repetition rate from all-fiber ring lasers to date.

The schematic diagram and the image of the laser oscillator are shown in Fig. 1(a) and (b), respectively. To shorten the cavity length as much as possible, compact devices should be used. Here we use an all-in-one device called polarization insensitive tap/iso/wavelength division multiplexer (PI-TIWDM) with active fiber pigtails to integrate the functions of 10% output coupler, isolator, and 980/1550 nm WDM. A 976-nm laser diode (LD, 1999CHB, 3SP Technologies) is used as the pump source for the oscillator. 24-cm highly Er-doped fiber (EDF, Er80-8/125, LIEKKI) with a core absorption of 80±8 dB/m at 1530 nm and an anomalous dispersion of -20 $ps^2$/km is used for providing gain in 1.5 μm for laser oscillation. The net dispersion in the cavity is calculated as -4800 $fs^2$. It is noted that no passive fiber is used in the laser cavity. The CNT-SA is prepared as follows: First, the commercialized high-pressure carbon monoxide (HiPco)-synthesized single-wall CNT powders (NanoIntegris) with diameters of 0.8–1.2 nm are fully dispersed into the N,N-dimethylformamide (DMF) solvent using an ultrasonic homogenizer (UH-600, SMT), then the CNT solution is directly sprayed onto the surface of the ferrule using a high-pressure spraying gun. At the same time, a dryer is used to evaporate the DMF solvent. After a certain time of spraying, the CNT film is formed on the fiber facet to act as the SA. Owing to the absorption of the gain fiber, the nonlinear transmission of the CNT-SA cannot be measured directly. We fabricate another CNT-SA on passive fiber under the same condition, and measure its saturable absorption property, shown in Fig. 1(c). The measurement setup and the fitting equation can be found in Ref. [29]. Results show that the CNT-SA has a modulation depth (MD) of 1.9% when peak intensity increases to 150 MW/$cm^2$, and the saturation intensity is measured as 22.8 MW/$cm^2$. The low saturation intensity is highly benefit for the low-threshold self-starting of the proposed mode-locked laser.

The optical output is directly measured by an optical spectrum analyzer (OSA, AQ6370D, YOKOGAWA) and an autocorrelator (FR-103XL, Femtochrome). The optical-to-electrical conversion is achieved by a high-speed photodetector with 25 GHz bandwidth (PD, 1414, NEWFOCUS). The converted pulse train is measured by an oscilloscope (DSA91304A, Keysight) and an electrical spectrum analyzer (ESA, RSA3045, RIGOL). The relative intensity noise (RIN) is analyzed by a RIN analyzer (PNA1, Thorlabs). The output power is recorded by a power meter (PM100USB, Thorlabs), and the stability of repetition rate is measured by a frequency counter (53181A, Agilent).

Turn-key, self-started mode-locking operation is achieved when the pump power reaches 108 mW. We slightly turn up the pump power to 150 mW while keeping its single-pulse mode-locking, the average output power is measured as 0.7 mW, and the output performance is plotted in Fig. 2. Figure 2(a) shows the optical spectra under a 0.1-nm resolution bandwidth (RBW), both in log scale and linear scale, where the latter one is well fitted by the sech$^2$ curve. The laser has a wide emission spectrum from 1545 nm to 1575 nm with a center wavelength $\lambda_c$ of 1562 nm and a full width at half maximum (FWHM) $\lambda_{FWHM}$ of 5.6 nm. No obvious Kelly sidebands are observed owing to the low peak power under such a high repetition rate. In Fig. 2(b), the corresponding autocorrelation (AC) trace is well fitted by the sech$^2$ curve, indicating the pulse is soliton shaped, which is expected to be obtained from an oscillator with anomalous dispersion. The pulse width is measured as 623 fs, and the time-bandwidth product (TBP) is calculated as 0.43, showing the pulse is slightly chirped. The pulse could be further compressed outside cavity. The laser owns an all-anomalous-dispersion cavity; shorter pulse width could be obtained if we properly manage the cavity dispersion.

Fig. 2(c) shows oscillogram of the laser pulse train with a roundtrip time of 1.3 ns. As shown in Fig. 2(d), the radiofrequency (RF) spectrum with a RBW of 300 Hz indicates the laser has a fundamental repetition rate $f_c$ of 783 MHz with a high signal-to-noise ratio (SNR) of 70 dB. The inset shows the higher-order harmonics up to 4 GHz with a 10-kHz RBW. In the broader RF spectrum, there is no other noisy frequency, showing the laser purely operates at the mode-locking regime. The intensity decline of the harmonic frequencies may result from the limited bandwidth of the ESA (4.5 GHz).

In our attempt for laser mode-locking, we find the mode-locking is highly dependent on the deposited CNT-SA. When thick CNT film (darker color in experimental observation) is deposited onto the ferrule, corresponding to a higher linear loss as well as high modulation depth [30], the laser operates in the Q-switching mode with higher possibility. When the laser operates at Q-switch regime, the CNT-SA is easy to be burnt owing to the high pulse energy. We gradually decrease the thickness of the CNT layer (decrease the modulation depth), and finally the mode-locking is achieved. Our finding is consistent with previous works reporting that small modulation depth of SA should be used for laser mode-locking in high repetition regimes [15, 31].

The relationship between the output power and pump power as well as the corresponding operation regimes is plotted in Fig. 3. In the process of increasing pump power, the laser experiences continuous-wave (CW) regime, Q-switching regime, and finally mode-locking regime when the pump power reaches 22 mW, 39 mW, and 108 mW, respectively. Under the pump power of 108 mW, the intracavity peak intensity is estimated around 25 MW/cm$^2$, the effective absorption of the CNT-SA is around 1% corresponding to Fig. 1(c). Once the pump power is higher than 230 mW, multi-pulse mode-locking is observed. Under the boundary state, the maximum output power under single-pulse mode-locking is measured as 1.21 mW. Although the output power is at mW level, it can be further amplified by an external amplification system.

We have noted that the threshold pump power is relatively higher than that of the CNT-mode-locked fiber lasers with lower repetition rates [28], the main reason may be the less net gain of the laser cavity and small modulation depth of the CNT-SA. The short active fiber needs more power to pump to reach the saturation intensity, and the loss induced by the coupling between fiber and TIWDM will cause an extra pump consumption. The small modulation depth would induce a higher self-starting threshold [32]. However, compared to lasers based on NPE scheme, the mode-locking pump power threshold is much lower. For examples, the thresholds are 650 mW in Ref. [22], 620 mW in Ref. [25], and 2 W in Ref. [23], respectively.

To evaluate the amplitude stability of the pulse train, the RIN of the free-running laser is measured. The RIN power spectral density (PSD) in the Fourier frequency range from 100 Hz to 1 MHz is shown in Fig. 4. The RIN-PSD maintains flat until 200 kHz, and then meets a hump located around 250 kHz, which could be the resonant relaxation oscillation frequency. The relaxation oscillation is caused by the finite upper-state lifetime of Er$^{3+}$ ions. Below the relaxation oscillation frequency of 250 kHz, the noise is introduced from amplified spontaneous emission (ASE) noise and pump noise. Above 250 kHz, the RIN-PSD drops dramatically and goes to the quantum noise limit [33]. Across the frequency range of [100 Hz–1 MHz], the integrated root mean square (rms) RIN is measured as 0.058%, which is comparable to that of the lasers with lower repetition rate [27], showing a good output stability. Due to the limitation of equipment, we are unable to measure the free-running timing jitter.

The free-running laser operates stably without any polarization controllers owing to the ultrashort cavity. The 6-hour operation stability of the laser is evaluated under room temperature (22±1 °C) without active protection stabilization. In the test period, seen from Fig. 5(a), there is no obvious change in the optical spectrum for both center wavelength and the border region, showing the robust mode-locking under free-running mode. We inject a high pump power and record the fluctuation of the laser's output power. As shown in Fig. 5(b), under the pump power of 200 mW, the laser's output power is measured as 1.01 mW, the total power variation is less than 14 µW in 6 hours, and the rms fluctuation is calculated as 0.22%. The periodic pulsation may result from the temperature change caused by the air conditioner. The repetition rate drift and the corresponding Allan deviation are shown in Fig. 5(c) and (d), respectively. A frequency drift of less than 2 kHz with a mean frequency of 783.019 MHz and a standard deviation (SD) of 514.653 Hz is obtained. The small jumping changes at around 2, 3 and 5.5 h in the frequency drift spectrum may be caused by environmental disturbance, such as opening/closing the door, or the vibration induced by people walking near the laser. The Allan deviation plot shows that, within the average time of 1000 s, the deviation is significantly decreased with a slope of $\tau^{-0.92}$. The minimum deviation of 107 mHz is achieved at 2000-s gate time. After this point, the deviation turns to increase, attributed to the environmental disturbance. Although the free-running laser shows good stability and robustness, we will further consider fully stabilizing it for subsequent high-precision applications.

In conclusion, we have demonstrated a compact and robust CNT-mode-locked Er-doped all-fiber laser delivering 623-fs pulses at center wavelength of 1562 nm, with a fundamental repetition rate up to 783 MHz. To the best of our knowledge, this is the highest fundamental repetition rate reported from all-fiber ring lasers. The all-active-fiber cavity design shortens the cavity length as much as possible, and a CNT-SA on fiber ferrule is adopted for laser mode-locking with low-threshold self-starting. The repetition rate can be boosted to over GHz using such a compact laser configuration by further shortening the fiber length and decreasing the device size.

**Acknowledgments**
This research is funded by Japan Society for the Promotion of Science (22H00209, 23H00174); and Core Research for Evolutional Science and Technology (JPMJCR1872). We also thank Mr. Hideru Sato and Mr. Raymond Chen for their kind personal donation to this research work.

**Disclosures**
The authors declare no conflicts of interest.

**Data availability**

Data underlying the results presented in this paper are not publicly available at this time but may be obtained from the authors upon reasonable request.

## References

1. T. Fortier and E. Baumann, Communications Physics **2**, 153 (2019).
2. S. Xing, D. M. B. Lesko, T. Umeki, A. J. Lind, N. Hoghooghi, T.-H. Wu, and S. A. Diddams, Apl Photonics **6**, 086110 (2021).
3. N. Picque and T. W. Haensch, Nature Photonics **13**, 146-157 (2019).
4. H. Kalayciog, P. Elahi, O. Akcaalan, and F. O. Ilday, IEEE Journal of Selected Topics in Quantum Electronics **24**, 8800312 (2018).
5. K. Sugioka and Y. Cheng, Light-Science & Applications **3**, e149 (2014).
6. N. Ji, J. C. Magee, and E. Betzig, Nature Methods **5**, 197-202 (2008).
7. J. Qin, R. Dai, Y. Li, Y. Meng, Y. Xu, S. Zhu, and F. Wang, Optics Express **26**, 25769-25777 (2018).
8. C. M. Wu and N. K. Dutta, IEEE Journal of Quantum Electronics **36**, 145-150 (2000).
9. Z. Zhao, L. Jin, S. Y. Set, and S. Yamashita, Optics Letters **46**, 3621-3624 (2021).
10. S.-S. Jyu, L.-G. Yang, C.-Y. Wong, C.-H. Yeh, C.-W. Chow, H.-K. Tsang, and Y. Lai, IEEE Photonics Journal **5**, 1502107 (2013).
11. X. Cao, J. Zhou, Z. Cheng, S. Li, and Y. Feng, Laser & Photonics Reviews **17**, 2300537 (2023).
12. R. V. Gumenyuk, D. A. Korobko, and I. O. Zolotovskii, Optics Letters **45**, 184-187 (2020).
13. A. Martinez and S. Yamashita, Applied Physics Letters **101**, 041118 (2012).
14. A. Martinez and S. Yamashita, Optics Express **19**, 6155-6163 (2011).
15. X. Gao, Z. Zhao, Z. Cong, G. Gao, A. Zhang, H. Guo, G. Yao, and Z. Liu, Optics Express **29**, 9021-9029 (2021).
16. G. Liu, X. Jiang, A. Wang, G. Chang, F. Kaertner, and Z. Zhang, Optics Express **26**, 26003-26008 (2018).
17. C. Li, Y. Ma, X. Gao, F. Niu, T. Jiang, A. Wang, and Z. Zhang, Applied Optics **54**, 8350-8353 (2015).
18. Y. Wang, R. Yang, Z. Chen, D. Pan, B. Luo, Z. Zhang, and J. Chen, *Frontiers in Optics + Laser Science 2023 (FiO, LS)*, Technical Digest Series (Optica Publishing Group, 2023), FD2.5.
19. R. Yang, M. Zhao, X. Jin, Q. Li, Z. Chen, A. Wang, and Z. Zhang, Optica **9**, 874-877 (2022).
20. S. Yamashita, Y. Inoue, K. Hsu, T. Kotake, H. Yaguchi, D. Tanaka, M. Jablonski, and S. Y. Set, IEEE Photonics Technology Letters **17**, 750-752 (2005).
21. R. Thapa, N. Dan, J. Zong, and A. Chavez-Pirson, Optics Letters **39**, 1418-1421 (2014).
22. T. Yang, H. Huang, X. Yuan, X. Wei, X. He, S. Mo, H. Deng, and Z. Yang, Applied Physics Express **6**, 052702 (2013).
23. J. Zhang, Z. Kong, Y. Liu, A. Wang, and Z. Zhang, Photonics Research **4**, 27-29 (2016).
24. Z. Chen, R. Yang, D. Pan, Y. Wang, B. Luo, J. Chen, and Z. Zhang, *Frontiers in Optics + Laser Science 2023 (FiO, LS)*, Technical Digest Series (Optica Publishing Group, 2023), JTu4A.70.
25. W. Du, H. Xia, H. Li, C. Liu, P. Wang, and Y. Liu, Applied Optics **56**, 2504-2509 (2017).
26. J. W. Nicholson and D. J. DiGiovanni, IEEE Photonics Technology Letters **20**, 2123-2125 (2008).
27. C. Kim, D. Kim, Y. Cheong, D. Kwon, S. Y. Choi, H. Jeong, S. J. Cha, J.-W. Lee, D.-I. Yeom, F. Rotermund, and J. Kim, Optics Express **23**, 26234-26242 (2015).
28. J. Sotor, G. Sobon, J. Jagiello, L. Lipinska, and K. M. Abramski, Journal of Applied Physics **117**, 133103 (2015).
29. M. Dai, B. Liu, G. Ye, T. Shirahata, Y. Ma, N. Yamaguchi, S. Y. Set, and S. Yamashita, Optics & Laser Technology **176**, 111002 (2024).
30. G. Sobon, A. Duzynska, M. Swiniarski, J. Judek, J. Sotor, and M. Zdrojek, Scientific Reports **7**, 45491 (2017).
31. Y. Ma, X. Zhu, L. Yang, M. Tong, R. A. Norwood, H. Wei, Y. Chu, H. Li, N. Dai, J. Peng, J. Li, and N. Peyghambarian, Optics Express **27**, 14487-14504 (2019).
32. S. Liu, Y. Chen, L. Huang, T. Cao, X. Qin, H. Ning, J. Yan, K. Hu, Z. Guo, and J. Peng, Optics Letters **46**, 2376-2379 (2021).
33. J. Kim and Y. Song, Advances in Optics and Photonics **8**, 465-540 (2016).

# Figures

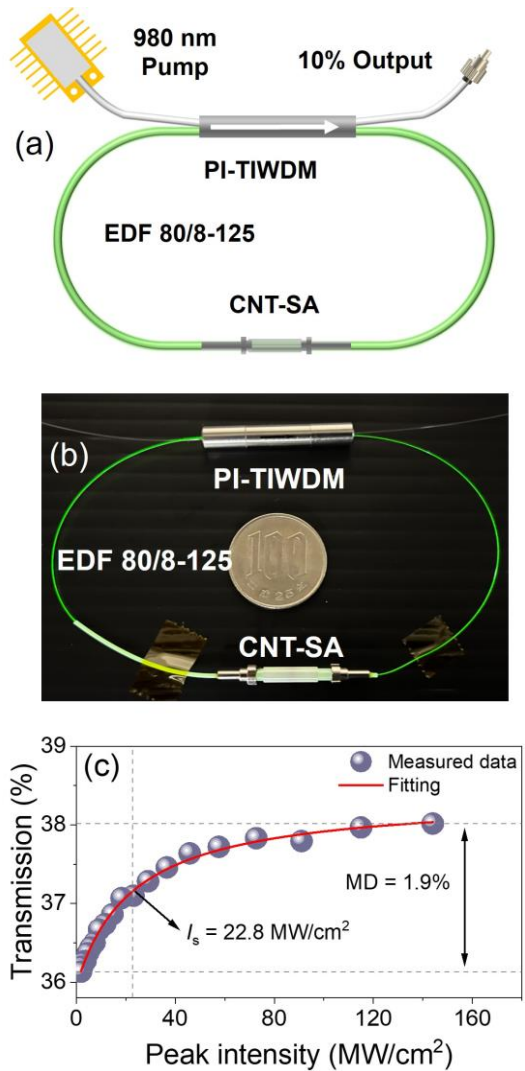

**Fig. 1.** (a) The schematic diagram of the proposed fiber laser; (b) The image of the proposed fiber laser; (c) Nonlinear transmission curve of the CNT-SA.

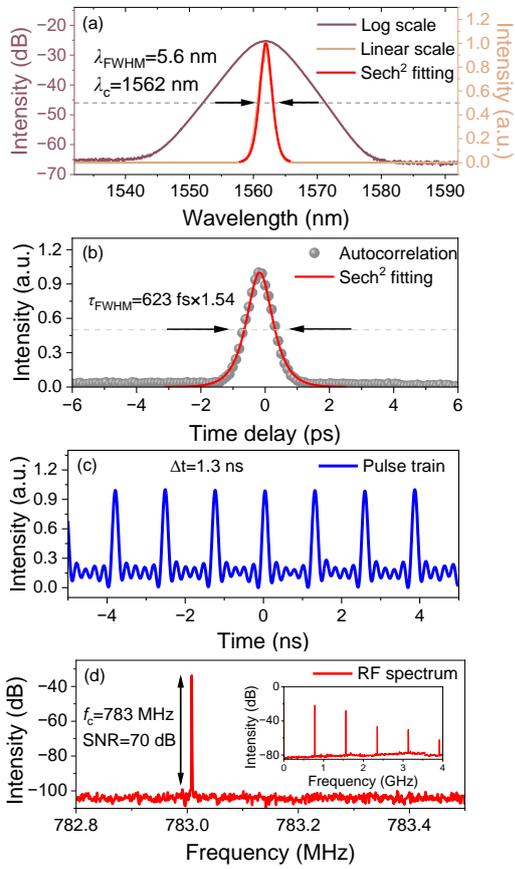

**Fig. 2.** Output performance of the proposed laser when the pump power is 150 mW: (a) Optical spectrum; (b) AC trace; (c) Pulse train and (d) RF spectrum.

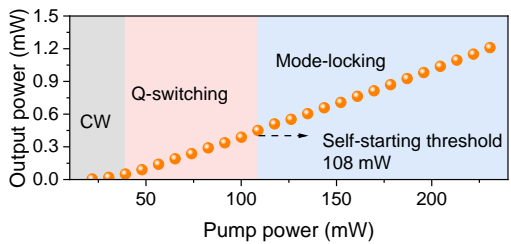

**Fig. 3.** The relationship between the output power with injected pump power and the corresponding operation regimes.

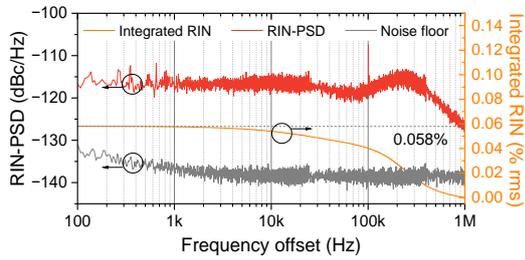

**Fig. 4.** RIN-PSD and the integrated rms RIN of the laser output in the Fourier frequency of [100 Hz–1 MHz].

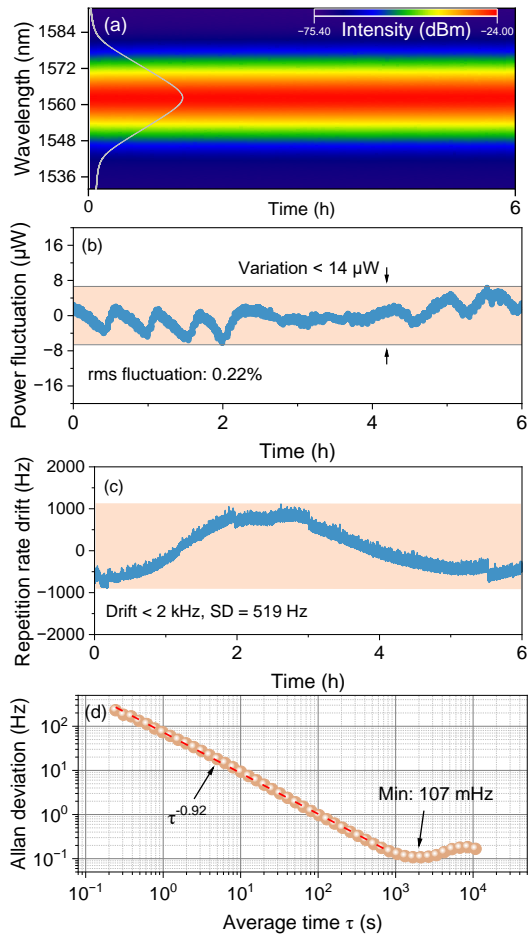

**Fig. 5.** Operation stability of the free-running fiber laser in 6 hours: (a) Optical spectrum; (b) Output power fluctuation under 200-mW pump power; (c) Repetition rate drift; (d) Allan deviation corresponding to (c).

**Table I.** Comparisons between high fundamental repetition rate Er-doped fiber ring lasers.

| Refs | Configuration | Fundamental repetition rate | Self-starting pump power | Mechanism | SNR (RBW) |
|---|---|---|---|---|---|
| [27] | All-fiber | 300 MHz | Not mentioned | CNT | 70 dB (1 kHz) |
| [28] | All-fiber | 358 MHz | 60 mW | CNT | 73 dB (1 Hz) |
| [25] | All-fiber | 374 MHz | 620 mW | NPE | 78 dB (1 kHz) |
| [26] | All-fiber | 447 MHz | N/A | CNT | N/A |
| [22] | All-fiber | 500 MHz | 650 mW | NPE | 85 dB (2 kHz) |
| [23] | Non-all-fiber | 517 MHz | 2 W | NPE | 50 dB (30 kHz) |
| [24] | Non-all-fiber | 895 MHz | N/A | NPE | 75 dB (30 kHz) |
| This work | All-fiber | 783 MHz | 108 mW | CNT | 70 dB (300 Hz) |